\documentstyle[twocolumn,prc,aps,epsfig]{revtex}
\input epsf
\newcommand{\be}{\begin{eqnarray}}
\newcommand{\ee}{\end{eqnarray}}
\def\lsim{\mathrel{\rlap{\lower3pt\hbox{\hskip1pt$\sim$}}
     \raise1pt\hbox{$<$}}} 
\def\gsim{\mathrel{\rlap{\lower3pt\hbox{\hskip1pt$\sim$}}
     \raise1pt\hbox{$>$}}} 

\begin{document}

\twocolumn[\hsize\textwidth\columnwidth\hsize\csname @twocolumnfalse\endcsname

\title{An Ultracold Strongly Coupled Gas: a Near-ideal Liquid}

\author {  Boris A. Gelman, Edward V. Shuryak and  Ismail Zahed}
\address { Department of Physics and Astronomy\\ State University of New York,
     Stony Brook, NY 11794-3800}

\maketitle
\begin{abstract}
Feshbach resonances of trapped ultracold alkali atoms allow to 
vary the atomic scattering length $a$. At 
very large values of $a$ the system enters an universal strongly 
coupled regime in which its properties---the ground 
state energy, pressure {\it etc.}---become independent of $a$. 
We discuss transport properties of such systems. In particular,
the universality arguments imply that the shear viscosity of 
ultracold Fermi atoms at the Feschbach resonance
is proportional to the particle number density $n$,
and the Plank constant $\hbar$: 
$\eta=\hbar n \alpha_\eta$, where $\alpha_\eta$ is a universal constant.
Using Heisenberg  uncertainty principle and Einstein's relation 
between diffusion and viscosity we argue that the viscosity
has the lower bound given by
$\alpha_{\eta} \leq (6\pi)^{-1}$. We relate the damping of
low-frequency density oscillations of ultracold optically trapped 
$^{6}Li$ atoms  to viscosity and find that the
value of the coefficient $\alpha_\eta$ is about $0.3$. 
We also show that such a small viscosity can not be explained by 
kinetic theory based on binary scattering.
We conclude that the system of ultracold atoms near the 
Feshbach resonance
is a near-ideal liquid.
\end{abstract}
\vspace{0.1in}
]
\begin{narrowtext}
\newpage
\section{Introduction}

A presence of the Feshbach resonance in binary collisions of alkali atoms
at ultracold temperatures provides an experimental control
of an effective strength of the inter-particle interaction. In the case of
ultracold dilute atomic systems the effective interaction strength is given
by an $s$-wave scattering length $a$. For a gas a  
``diluteness parameter'' $n\, a^3$, where $n$ is the number density,
is much less then unity.\footnote{A typical density of ultracold trapped
atomic gases is \\ $n \sim 10^{13}-10^{15} \, cm^{-3}$.}
An atomic system can be tuned to a
regime of the Feshbach resonance using an external magnetic field. On
resonance the scattering length and ``diluteness parameter''
formally diverge. The system can no longer be described as a dilute
weakly interacting Fermi or Bose gas. This regime is often referred to as
strongly coupled regime. 

As external magnetic field is tuned to a critical value
corresponding to the Feshbach resonance the effective dimensionless
coupling constant describing two-body interaction formally diverges.
In the case of ultracold alkali atoms this coupling constant $g$ is given
in terms of the $s$-wave scattering length $a$ by
$g=4 \pi \hbar^{2} a /m$, where $m$ is the mass of an atom.
Since the coupling constant $g$ goes to infinity at the Feshbach resonance
it is natural to expect that the thermodynamic properties of the atomic system
become independent of $g$ and consequently of the nature of the two-body
interaction \cite{HH}.
Instead, these properties are given in terms of the {\em universal constants} 
times the appropriate dimensional factors which for ultracold atoms are
given in terms of the remaining parameters, namely, the Plank constant
$\hbar$, particle density $n$, and an atomic mass $m$.  

For example, the pressure $p(n)$ of ultracold 
atoms at the Feshbach resonance is proportional to that of the 
ideal Fermi gas, $p_0 (n)\sim  n^{5/3}/m$,
\be \label{EoS_atoms}
{p(n)\over p_0 (n)}= 1-\beta 
\ee
where $\beta$ is an universal dimensionless constant.
For ultracold fermionic atoms $\beta\approx 0.5$.
The ``universality" principle means that the value of  $\beta$ is the same for
any strongly coupled system of fermions, such as cold atoms or
dilute neutron matter. This constant is expected to depend on
the particle spin. 

An atomic system near the Feshbach resonance is an example of a quantum
many-body system with a large or formally infinite 
dimensionless parameter describing two-body interaction. 
Other examples include: 
a strongly coupled quark-gluon plasma (sQGP) \cite{sQGP},
a strongly coupled classical electromagnetic plasma and 
a strongly coupled ${\cal N}=4$ supersymmetric 
Yang-Mills gauge theory.
These systems are briefly discussed in section \ref{sec_discussion}.

In all of the above systems one finds that the difference between weak 
and strong coupling regimes in the total  (free) energy, although
nonzero, is nevertheless much less dramatic than the corresponding
difference in  their transport 
parameters---viscosity, diffusion, {\it etc.}---
which change by orders of magnitude. In these cases the strongly coupled
matter behaves as a liquid, with the the mean free path comparable 
or even smaller than the inter-particle distance. 

The goal of this paper is to show that 
the {\em transport properties} of a system of ultracold atoms 
change dramatically as the system is driven from
a weakly interacting regime to the regime of the Feshbach resonance. 
As we will discuss in details this change is manifested in a steep
decrease by about two orders of magnitude of the damping rate 
of the collective oscillations of a cloud of $^{6}Li$ atoms
near the Feshbach resonance observed in the experiments of
Kinast {\it et al.} and Bartenstein {\it et al.}.
This decrease cannot be attributed to the phase transition that the system 
is expected to undergo at a low enough temperature.
The maximum of the oscillations is near the surface of the atomic cloud
where the temperature is above the local phase transition temperature.
Furthermore, the transport coefficients such as viscosity and diffusion 
display universal properties near the Feshbach 
similar to the pressure and energy density.

\section{The shear viscosity of ultracold fermionic atoms from oscillation
damping}
\label{sec_visc}

In this section we use the equations of hydrodynamics to describe
the damping rate of the collective excitations of the ultracold Fermi atoms
near the Feshbach resonance.
A spectacular manifestation of the hydrodynamical behavior of ultracold
optically trapped atomic clouds near the Feshbach resonance
is the {\it elliptic flow} observed after
the trap is switched off \cite{elliptic_flow,elliptic_flow_two}. 
Such flow is also 
observed at higher temperatures. Thus it is not by itself
related to quantum low-temperature phenomena like condensation or 
superfluidity but rather it is a manifestation of a strongly coupled regime.
A good quantitative description of the elliptic flow
was obtained in a framework of 
ideal hydrodynamics ({\it i.e.} neglecting viscosity).
In principle, this comparison by itself may provide an upper limit on 
the viscosity.

Additional evidence for the hydrodynamical behavior is provided by
the experiments of Kinast {\it et al.} \cite{duke,duke2} and 
Bartenstein {\it et al.} \cite{osc_omega_damp} on collective oscillation
of ultracold trapped $^6\!Li$ atoms in the vicinity
of the Feshbach resonance. The measured frequencies of the lowest 
modes are in a good agreement with the predictions of ideal non-suprefluid
hydrodynamics.

Our main point is that the damping of these collective excitations
can be described by including the dissipative (in particular, 
{\em viscous}) terms in the hydrodynamic equations.
As will be shown in the next section, the viscous terms are very small
at least for the low frequency collective excitations. Thus, these terms can 
be included perturbatively.

If the collision rate of atoms is large enough to establish a local equilibrium
the collective vibrations of the atomic cloud can be described using 
standard hydrodynamical theory \cite{L&L}. The collective
vibrations are described by the local density $n(\vec{r})$, pressure 
$p(\vec{r})$ and velocity $\vec{v}(\vec{r})$ 
which are the solutions of the continuity
equation, Euler equations of motions and the equation of state, 
respectively,
\be 
& m &\, {\partial n\over \partial t} + \nabla \cdot \left(m\, n\, \vec{v}
\right)
=0 \, , \nonumber \\
& m & \,n\, {\partial \vec{v}\over \partial t} + m\, n\, 
\left(\vec{v} \cdot \nabla\right) \, \vec{v} = -\nabla p -n\, \nabla V \, ,
\nonumber \\
& p & =A \, n^{\gamma +1} \, ,
\label{hydro}
\ee
where $V=(1/2)\, m \, \sum_i \omega^{2}_{i}\, r^{2}_{i}$ is the harmonic
trapping potential, $A$ is a constant, and $\gamma$ is the 
polytropic index. 
As discussed in \cite{HH} a cold gas of strongly interacting fermions at the
has a universal equation with polytropic index $\gamma=2/3$.

The lowest collective modes correspond to the small vibrations of the density,
pressure and velocity around their equilibrium values, 
$n_{eq}$, $p_{eq}$ and $\vec{v}_{eq} =0$. These values are
determined by the static limit of the Eqs.~(\ref{hydro}) in which the Euler
equation takes the form $\nabla p_{eq} + n_{eq} \nabla V =0$. The equilibrium
density is
\be n_{eq}(r)= n_{eq} (0) 
\left(1-\sum^{3}_{i=1} {r^{2}_{i} \over R^{2}_{i}} \right)^{1/\gamma} \, ,
\label{neq}
\ee
for $\vec{r}$ inside the ellipsoid 
\be
{x^2\over R^{2}_{x}} + {y^2 \over R^{2}_{y}}+ {z^2 \over R^{2}_{z}}=1 \, 
\label{ellips}
\ee 
and zero outside. In equation~(\ref{neq}) $n_{eq} (0)$ 
is the equilibrium density of the atomic cloud at the center
of the harmonic trap ($\vec r =0$) and 
\be
R_i = \sqrt{ {p_{eq} (0) \over n_{eq} (0)} \, 
{2 (\gamma+1) \over \gamma \, m \omega^{2}_{i}}} \,
\ee
are the  radii of the cloud with $p_{eq} (0)$ being the equilibrium
central value of the pressure. These radii can be expressed in terms of the
chemical potential $\mu$. Using the Gibbs-Duham relation, 
$dp = n \, d \mu$ (valid at constant temperature) and the equation of state,
$p = A n^{\gamma +1}$, one can easily show that
\begin{equation}
{p\over n} = {\gamma \over \gamma +1} \, \mu
\label{p_over_n}
\end{equation}
so that the Thomas-Fermi radii are
\be
R_i = \sqrt{2 \, \mu \over m \omega^{2}_{i}} \, ,
\label{Rimu}
\ee
independent of $\gamma$. 

The chemical potential $\mu$ in Eqs.~(\ref{p_over_n}) and (\ref{Rimu}) 
includes interactions of the trapped atoms in the vicinity of the Feshbach
resonance. The effect of the interaction in the unitary
limit is to scale the chemical potential of the non-interacting trapped Fermi
gas \cite{elliptic_flow,mu_strong}:
\begin{equation}
\mu = \mu^{(0)} \, \sqrt{1-\beta} \, ,
\end{equation}
where $\beta \approx 0.5$ is the universal constant discussed in the
introduction and $\mu^{(0)}$ is the chemical potential of 
$N$ spin-$1/2$ noninteracting 
fermions confined in a harmonic trap \cite{P&S},
\be
\mu^{(0)}= \hbar \bar{\omega} \left(3 N\right)^{1/3} = k_B \, T_F
\, ,
\label{mu}
\ee
where $\bar{\omega}^{3}=\omega_1 \, \omega_{2} \, \omega_{3}$ and
$k_B$ is the Boltzmann constant. The last equation
also defines the Fermi temperature $T_{F}$ of the system.

Linearizing Eqs.~(\ref{hydro}) to describe
small density $n=n_{eq} + \delta n \, e^{i \omega t}$
and velocity $\vec{v} \, e^{i \omega t}$ oscillations 
one can obtain the following equations for the density and 
velocity amplitudes:
\be -m\omega^2 \delta n & = & 
\nabla \cdot \left(n_{eq}\nabla\left({1\over n_{eq}}
{ dp \over dn} \delta n\right) \right) \, \nonumber \\ 
-\omega^2 \vec{v} & = & \nabla \left(\vec{v} \cdot \nabla V \right) +
\gamma \, \nabla \cdot \vec{v} \, \nabla V \, .
\label{modes}
\ee

The form of the collective vibrational modes which are the 
solutions of Eq.~(\ref{modes}) depend on the symmetry of the trap potential.
If the confining potential is isotropic,  
$\omega_{1}=\omega_{2}=\omega_{3}$, the collective modes have spherical 
symmetry and can be characterized by definite angular momentum and its $z$ 
component, {\it l} and {\it m}. The monopole mode, ${\it l}={\it m}=0$, 
has a velocity profile which is proportional to $\vec{r}$. Such modes
are referred to as breathing modes. The dipole mode ${\it l}=1$ involves
the motion of the center of mass of the cloud and is not usually excited in
the experiments we are interested in. There are five degenerate quadrupole 
modes corresponding to ${\it l}=2$ with ${\it m}=0, \pm 1, \pm2$.

We will discuss the collective modes that are excited in an 
axially symmetric trap with $\omega_{1}=\omega_{2}=\omega_{r}$ and 
$\omega_{3}=\omega_{z}=\lambda \omega_{r}$, 
where $\lambda$ is a constant. In the experiment of Bartenstein {\it et al.}
\cite{osc_omega_damp} 
$\lambda$ is $0.03$\footnote{ Such traps with $\lambda \ll 1$ are
referred to as cigar-shaped or prolate.}. In such traps the collective modes
corresponding to different angular momenta but the 
same $z$-component {\it m} are mixed. 

To find these modes we look for the solutions of Eq.~(\ref{modes})
in the form\footnote{Such flow with velocity components proportional 
to the position is often referred to as Hubble flow.} 
\be
\vec{v}= (a_x x, a_y y, a_z z) = (a_r x, a_r y, a_z z) \, .
\label{vHubble}
\ee
The set of equations in Eq.~(\ref{modes}) for 
$\vec{v}$ reduces to the secular equation for the eigenfrequencies
and the corresponding eigenvectors. The three frequencies are
\cite{hh_modes},
\be
\omega^{2}_{1,2} 
& = &\omega^{2}_{r} (1+\gamma +{1\over 2}(\gamma+1) \lambda^2
\nonumber \\
& \pm &
{1\over 2}\, \sqrt{(\gamma+2)^2 \lambda^4+(\gamma^2-3\gamma-2) \lambda^2+
4 (\gamma+1)^2} ) \, ,
\nonumber \\
\omega_{3} & = & \sqrt{2} \, \omega_r \, ,
\ee
where in the first equation the plus and minus signs correspond to axial and
radial modes with frequencies 
$\Omega_{z} = \omega_{1}$ and $\Omega_{r} = \omega_{2}$ respectively. The
third frequency (which is the same as one of the frequencies in the case 
of the spherical trap) is the frequency of the two remaining degenerate modes
with ${\it l}=2$, ${\it m}=\pm 1\, \pm 2$. For the cigar-shaped traps, 
{\it i.e.} in the limit of a very small $\lambda$, the axial and radial 
frequencies reduce to
\be 
{\Omega_z \over \omega_z} & = & \sqrt{3-{1\over \gamma+1}} \, , 
\nonumber \\
{\Omega_r \over \omega_r} & = &  \sqrt{2\gamma+2} \, .
\label{omegas}
\ee
For the equation of state with the polytropic index $\gamma=2/3$.
the axial and radial frequencies are, respectively,
\be 
{\Omega_{z} \over \omega_{z}} & = & \sqrt{12\over 5} =1.549 \, , 
\nonumber \\
{\Omega_{r} \over \omega_{r}} & = & \sqrt{10\over 3} =1.826 \, .
\label{omegas_values}
\ee

The corresponding velocities, Eq.~(\ref{vHubble}), in the limit as 
$\lambda \to 0$ are given
for the axial mode by, 
\be
{a_z \over a_r}  = - {2 (1+\gamma) \over \gamma}=-5\,,
\label{azaraxial}
\ee
and for the radial mode by,
\be
{a_z \over a_r} & = & {\gamma\over \gamma+1}\, \lambda^2
={2\over5}\, \lambda^2 \, , 
\label{azar}
\ee
where $\gamma= 2/3$ was used.

The frequency of the axial mode in Eq.~(\ref{omegas_values}), 
agrees very well with the value of approximately $1.55$ 
measured by Bartenstein {\it et al.} \cite{osc_omega_damp}
The frequency of the radial mode at the Feshbach resonance
measured by Kinast {\it et al.} 
($\Omega_{r} / \omega_{r} =1.829$) also agrees remarkably well 
with the value given in
Eq.~(\ref{omegas_values}) \cite{duke2}.

However, the frequency of the radial mode measured in the experiment by 
Bartenstein {\it et al.} (\cite{osc_omega_damp}) is $1.67$ which is about
$20\%$ below the corresponding value in Eq.~(\ref{omegas_values}).
This deviation complements a sudden jump in the frequency and 
especially in the 
damping rate of the radial mode \cite{osc_omega_damp} not far from 
the resonance on the ``BCS'' side.  There is no jump
at this particular value of the scattering length in the 
axial mode, so this phenomenon cannot possibly be associated with a change
in the equation of state  or transport properties by itself. Thus, an 
extra source of dissipation in the radial mode in the experiment by
Bartenstein {\it et al.} must thus be associated with a non 
hydrodynamical effects\footnote{We are grateful to R.~Grimm
for a discussion about a possible source of this effect.}. 

We now apply viscous hydrodynamics to describe the damping of the
collective modes in the vicinity of the Feshbach resonance. The
primary source of dissipation in the hydrodynamic limit is shear
viscous flow \cite{KPS}. The rate of change of the total
energy of a mode is given by \cite{L&L}
\be 
\left<{d E \over d t}\right> & = &
-\int \, {\eta \over 2} \, 
\left({\partial v_i \over \partial x_k} + 
{\partial v_k \over \partial x_i}-{2\over 3} \, \delta_{ik} \,
\nabla \cdot \vec{v} \right)^2 \, d^3 r \nonumber \\
& = & -\, {2\over 3} \, a^{2}_{r} \, \left(1-{a_{z} \over a_{r}} \right)^2 \, 
\int \, \eta(r) \, d^3 r \,,
\label{rate}
\ee
where $\eta$ is the coefficient of the shear viscosity and the ratio 
$a_{z}/a_{r}$ is given in Eqs.~(\ref{azar}) for each mode.

A damping rate of a collective mode can now be obtained by dividing the
rate of change of energy, Eq.~(\ref{rate}) by half of the the time-averaged
total energy of the vibrational mode which is equal to maximum
kinetic energy of the mode
\be
\left<E\right>  & = &  {m\over 2}\, \int n_{eq} (\vec{r})\, v^2 (\vec{r})\,
d^3 r \nonumber \\
& = & 
{ \pi^2 \over 128} \, m\, n_{eq}(0)  \, a^{2}_{r} \, R_x\, R_y \, R_z \, 
\left(R^{2}_{x} +R^{2}_{y} + {a^{2}_{z} \over a^{2}_{r}} \, R^{2}_{z} \right) 
\nonumber \\
& =  & {1\over 16} \, m\, N a^{2}_{r} \, 
\left(2 \, R^{2}_{r} + {a^{2}_{z} \over a^{2}_{r}} \, R^{2}_{z} \right) \,,
\label{totalE}
\ee
where Eq.~(\ref{neq}) was used with $\gamma=2/3$. 
In the last step the product of Thomas-Fermi radii was expressed in terms 
of the total number of particles $N$ in the cloud,
\be
N & = & \int\, n_{eq}(\vec{r}) \, d^3 r =
\int \,n_{eq}(0)\,
\left(1-\sum^{3}_{i=1} {r^{2}_{i} \over R^{2}_{i}} \right)^{3/2}
\, d^3 r 
\nonumber \\
& = & {\pi^2 \over 8}\,n_{eq}(0)\,  R_x \, R_y \, R_z \,,
\ee
and $R_{x}=R_{y}=R_{r}$ for the axially symmetric trap.

Using Eqs.~(\ref{rate}) and (\ref{totalE}) the damping rate is
\be 
\Gamma & = & {1\over 2}
\left| {\left<d E/d t\right> \over 
\left<E\right>} \right| 
\nonumber \\
& = & {16\over 3\, m\, N} \,
{\left (1-{a_z \over a_r}\right)^2 \over
\left(2\, R^{2}_{r} + {a^{2}_{z} 
\over a^{2}_{r}} \, R^{2}_{z} \right)} \,  \int \, \eta(r) \, d^3 r \, .
\label{Gamma}
\ee
This expression will be used in the next section to extract the
coefficient of shear viscosity of the strongly coupled
atoms near the Feshbach resonance.

\section{ Quantum viscosity   }
\label{sec_uni}
\subsection{Generalities}

In the weak coupling regime of a small scattering length $a$, 
transport properties
are determined by the particle mean-free path 
\be 
\label{mfp}
l_{\rm mfp}={1\over n\sigma} \, ,
\hspace{1cm} \sigma=4\pi a^2 \,\,.
\ee

In the strong coupling regime, when $a\rightarrow \pm \infty$,
Eq.~(\ref{mfp}) becomes meaningless. Since
the cross section does not diverge but is bounded by unitarity,
this regime is sometimes referred to as ``unitarity limited'' one.
Indeed, the maximal possible two-body cross section is limited by 
 $\sigma_{\rm max}=4\pi/k^2$ for fixed collision energy ($k$ is
the wavenumber of relative motion).
Thus, one may think that the mean free path is actually
 $l_{min}=1/n\sigma_{\rm max}$. However, this is  too naive as well, since 
at strong coupling there is no reason to limit kinetics to a picture of
propagating particles rarely suffering only binary collisions.

Whatever the microscopic structure of matter may be, the {\em viscous
hydrodynamics} is an adequate description of low frequency dynamics.
It is important that it is based
on expansion in $inverse$ powers of the cross sections,
or expansion in small mean free paths $l$, so 
the stronger the interaction the better this approach is expected to work.
The viscosity is in general defined via the dissipative part of the stress
tensor and can be defined without any assumption on the underlying matter. 
It appears in observables like the sound dispersion law
\be 
\omega=c_s k+{i\over 2}{4\eta \over 3 m n} k^2 
\ee
and thus can be measured. In fact, the damping rate in eq.~(\ref{Gamma})
is the corresponding analog for trapped atoms.

\subsection{Universality of transport coefficients}

For simplicity we discuss mostly the zero temperature limit in this
section. The temperatures in the experiments of
of Kinast {\it et al.} \cite{duke,duke2} and 
Bartenstein {\it et al.} \cite{osc_omega_damp}
range from $0.1 \, T_F$ to $0.03 \, T_F$. 

One of the basics of low temperature
physics of weakly coupled systems is that at low $T$  
the mean free path of quasiparticles goes to infinity,
together with viscosity. There are multiple examples of such behavior,
e.g. in liquid $^{4}{\it  He}$ the minimal viscosity is around $T_c$ and then
it rises as $T\to 0$. The normal component at low $T$ is
thus very viscous.

However, this picture is no longer true at weak coupling.
Strongly coupled normal component is also expected to have
low viscosity
even at low $T$ \footnote{For clarity, we remind again that
experimentally observed oscillations are not an elementary
quantum phonon state, but a highly excited state to which
we apply the viscous hydrodynamics.}. 
Furthermore, we suggest that such oscillations even at
$T\rightarrow 0$ would have a damping described by a 
``quantum viscosity'' proportional to the Plank constant,
reflecting limitation on particle localization even at zero
temperature. Indeed, at zero temperature a single interaction 
parameter---the scattering length $a$---diverges at the Feshbach resonance. 
The only remaining length scale is given by the inter-particle distance, 
$n^{-1/3}$. Thus, the mean free path must be of this scale.

These arguments lead to universal relations for transport coefficients.
In particular, the only form of the viscosity is
\be 
\label{alpha_eta}
\eta=n\hbar \alpha_\eta
\ee
where an universal dimensionless coefficient $\alpha_\eta$
is introduced. Note that it is independent on the particle mass.
The same arguments are also valid for strongly
coupled bosonic atoms, although of course with a different $\alpha_\eta$.

At nonzero temperature viscosity has a similar form,
 \be 
\label{beta_eta}
\eta=n\hbar \alpha_\eta + s(T) \hbar \beta_\eta
\label{visc_temp} 
\ee
where the second temperature-dependent term 
contains the entropy density and another
universal coefficient $\beta_{\eta}$. 
This second term reproduces the finite temperature.
dependence of the damping observed by the Duke group, although
we do not have any other justification for this form of this term. 
We repeat, that none of these terms are reproduced
by binary collisions, neither in magnitude nor in parametric dependences.

Similarly, the characteristic time scale
in strongly coupled liquid  is given by the Fermi energy
and the Plank constant $\hbar$,
$\tau^{-1}\sim \epsilon_F/\hbar\sim \hbar n^{2/3}/m$. 
even if it include purely Bosonic atoms.
Thus, the scattering rate should 
be simply proportional to this scale with another universal
coefficients, although different for bosons and fermions.

The next question is what are the values of these dimensionless
parameters. For an ordinary liquid, these constants are of order one.
For a quantum liquid, such as $^{4}He$ with a large quasiparticle
mean free path the value of $\alpha_{\eta}$ exceeds that of ordinary
liquids by about three orders of magnitude.

If the dimensionless parameters are much smaller than one,
we deal with exotic near-ideal liquid.
Although the available data available is not conclusive. It is 
is quite possible that the ultracold fermionic
atoms near the Feshbach resonance are in the regime of near-ideal
liquid.

\subsection{Experimental estimate for universal viscosity coefficient}

It was shown in section \ref{sec_visc} that
the damping rate of the small collective oscillations
is proportional to the volume-integrated viscosity, Eq.~(\ref{Gamma}).
The universality relation Eq.~(\ref{alpha_eta})
reduces the integral in Eq.~(\ref{Gamma}) 
to the total number of particles times 
the coefficient $\alpha_\eta$ which we want to determine. 
Thus, the damping rate is
\footnote{The universal relation should not be valid  
near the edges of the system, at less than one mean-free-path or 
at optical depth less than one, where dissipation is larger.
However, the edge includes only about one percent
of particles and their contribution can be neglected.},
\be
\Gamma = {16\over 3\, m} \,
{\left (1-{a_z \over a_r}\right)^2 \, \hbar \, \alpha_{\eta} \over
\left(2\, \lambda^{2} + {a^{2}_{z} 
\over a^{2}_{r}} \, R^{2}_{z} \right)} \, .
\label{Gamma2}
\ee

For the axial mode in the cigar-shaped potential trap using 
the ratio $a_{z}/a_{r}$ from Eq.~(\ref{azaraxial}), 
the coefficient $\alpha_{\eta}$ is
\be
\alpha_\eta \approx {25 \, m \, R^{2}_{z}\, \Gamma_z \over 192 \hbar} 
\ee
The experimental values are $\Gamma_z/\omega_z=0.0036 $ 
at the point of the Feshbach resonance, with the minimal
value of $(\Gamma_z/\omega_z)|_{min}=0.0015 $ slightly off the resonance
\cite{osc_omega_damp}.
Using the axial trap frequency
$\omega_z \approx 140\, Hz$
we get our final result for the minimal dimensionless
viscosity\footnote{The error  is comparable to the value itself, as
can be seen from experimental data points. Ironically, the situation
with dimensionless viscosity of quark-gluon plasma is quite similar.}   
\be 
\label{eqn_minvalue}
\alpha^{(z)}_{\eta}|_{min} \approx 0.3  \,.
\label{etaz}
\ee
The fact that we get a number of order unity is an indication
that the suggested ``quantum viscosity'' indeed exists and is
described by the universality arguments. 

For the radial mode, using Eq.~(\ref{azar}) and the fact 
that $R_{z}=R_{r}/\lambda$ we get,
\be
\alpha_\eta \approx {3 m \, R^{2}_{x}\, \Gamma_r \over 8 \hbar}\,. 
\ee

In the experiment of Bartenstein {\it et al.} 
\cite{osc_omega_damp} with the radial trap frequency 
$\omega_{r} \approx 4700 \, Hz$ the damping rate of the radial
mode at the Feshbach resonance is  $\Gamma_r=0.0625 \, \omega_r$.
Thus, the coefficient $\alpha_{\eta}$ extracted from the damping rate of
the radial is
\be 
\alpha^{(r)}_{\eta} \approx 1.1  \,.
\ee

The coefficient $\alpha_{\eta}$ extracted from the damping rate of
the radial mode measured in the experiment of Kinast  {\it et al.}
\cite{duke2} is
\be 
\alpha^{(r)}_{\eta} \approx 0.2  \, .
\ee 

The damping rate of the radial mode measured by
Kinast  {\it et al.} \cite{duke2} at temperature comparable to the
Fermi energy is more or less consistent with an estimate from
kinetic theory discussed in the next section. 

Kinast {\it et al.} \cite{duke2} measured the temperature
dependence of the damping rate. 
As the temperature is lowered the damping rate
${\em decreases}$ in contradiction to the kinetic estimates 
in section \ref{sec_kinetic}, but in agreement with our reasoning. 
Unfortunately, the temperature is not 
small enough to conclude whether the universal 
``quantum viscosity'' regime at $T=0$
is reached and $\alpha_{\eta} \neq 0$ or not, 
and whether the same universal viscosity is responsible for damping of both 
axial and radial modes. The second term in Eq.~(\ref{visc_temp})
proportional to the entropy density can explain this data if value of
$\beta_{\eta}$ is of order unity.

\section{Comparison with the traditional kinetic theory}
\label{sec_kinetic}

In this section we will show that 
the kinetic theory based on the
notion of binary collisions fails
to describe the damping of the axial mode discussed above.

Let us start with an order of magnitude estimates of the collision
rates and viscosity neglecting
Pauli blocking and using the largest
 ``unitary limited'' cross section $\sigma=4\pi/k_F^2$.
The collision rate at the center of the trap estimated like this gives
\be \tau^{-1}_{coll}=n(0)\sigma v_f \sim  10^5 s^{-1} \ee
where the last number corresponds to conditions of the experiment of
Bartenstein {\it et al.} \cite{osc_omega_damp}. Compared
with oscillation frequencies, $ \omega_r=4712\, Hz,\omega_z=141.9 Hz$
of the trap, which leads to a conclusion that only the latter mode have a
chance to be hydrodynamical.

The mean free path, ${\it l_{\rm mfp}}$,
of a particle is of order $1/(n \sigma)$,
while the
shear viscosity is
\be
\eta  \sim m \, \bar{v} \, n \, {\it l_{\rm mfp}} = {m \, v \over \sigma}
\, ,
\ee 
where $v$ is the average velocity of a particle.
In the limit of zero temperature the velocity is set by the Fermi momentum,
$m\, v_{F} = \hbar \, k_{F}$. In the vicinity of the Feshbach resonance
the cross section is unitary bounded, $\sigma <\sigma_{max}=4 \pi/k^{2}_{F}$. 
So, if we take its $maximal$ value  (and still ignore
for a moment  Pauli blocking), we will get a $minimal$ viscosity
which may follow from binary collisions: 
\be
{\eta \over \hbar \, n} >  {40 \over 6 \pi} \, .
\ee 
This inequality is strongly violated in experiment, as shown above:
this ``minimal binary'' value is in fact larger than observed value, and 
forty times larger than the derived bound possible for a liquid.

Furthermore, if the
fermionic atoms were in the Fermi liquid regime,
the collision rate will be significantly
lowered by Pauli blocking which should lead to a suppression factor about 
$\left(T /T_F\right)^2\sim 1/1000$
in the experimental conditions. If true, the
oscillations then would be basically collisionless and no
hydro phenomena would be present.
In a picture of BCS-type pairing, with relatively small
modification of Fermi sphere, $T$ in the above formula
to  be substituted by a gap, so the rescattering suppression
would be of the order of $ ({\Delta/ \epsilon_F})^2$. The gap value is 
not well known, but this suppression factor is still  about
1/100 or so. We must then conclude that both pictures are wrong
and in fact there seem to be no Pauli blocking
whatsoever\footnote{After observation of elliptic flow this issue was
discussed in literature and the MIT group \cite{MIT_collisions}
have argued that this might have been due to strong deformation
of Fermi sphere in exploding gas. This explanation obviously would not
be applicable to small amplitude oscillation we study here.}.

To make more quantitative conclusion we will derive here the damping
rate of a collective mode in a axially symmetric trap applying the
traditional kinetic equation to an almost ideal Fermi gas with 
unitary limited cross section.  

A damping rate of a collective mode in the kinetic theory is determined
by a relaxation time which is a measure of how fast a particle distribution
function $n (\vec{p}, \vec{r}, t)$ for a given collective mode
takes an equilibrium form. Both the time dependent and equilibrium
distribution functions are the solutions of the kinetic equation. 
The equilibrium distribution for a Fermi gas is
\be
n (\epsilon, \vec{r}) = 
\left(e^{(\epsilon - \mu + V(r)) /k_B \, T} +1 \right)^{-1} \,,
\ee 
where $\epsilon = p^2/ 2 m$. 

During an oscillation the distribution function is different from the 
equilibrium one. The collisions between particles cause the non-equilibrium
distribution function to ``relax'' to the equilibrium form. These collisions
are the source of the damping of the oscillations. 

As shown in \cite{P&S} the damping rate of the oscillations of Fermi gases
is equal to 
\be
\Gamma = { \left<\left(p^{2}_{1,z} - {p^{2}_{1}\over3} \right) 
\Gamma \left[p^{2}_{1,z} - {p^{2}_{1}\over3} \right]\right> \over
\left<p^{2}_{1,z} - {p^{2}_{1}\over3} \right> } \, ,
\ee 
where
\be
& &  \left<\left(p^{2}_{1,z} - {p^{2}_{1}\over3} \right) 
\Gamma \left[p^{2}_{1,z} - {p^{2}_{1}\over3} \right]\right> = 
\nonumber \\
& & {1 \over 4 (2 \pi \hbar)^6}
\int \, d^{3} r \, d^{3} p_{1} \, d^{3} p_{2}\, d^{3} p^{\prime}_{1} 
\, d^{3} p^{\prime}_{2} \, \left(\Delta \Phi\right)^2 \, 
\nonumber \\
& & W \,
\delta^{3} (\vec{p}_1 +\vec{p}_2-\vec{p}^{\prime}_1-\vec{p}^{\prime}_2) \,
\delta (\epsilon_1 +\epsilon_2-\epsilon^{\prime}_1-\epsilon^{\prime}_2) \,
\nonumber \\
& & n_{1} \, n_{2} \, \left(1-n^{\prime}_{1} \right) \,  
\left(1-n^{\prime}_{2} \right) \, ,
\label{N}
\ee
and
\be
\left<\left(p^{2}_{z} - {p^{2}\over3} \right)^2\right> = 
\int \, d^{3} r \, d^{3} p \, \left(\Phi\right)^2 \, 
n \, \left(1-n \right) \,  ,
\label{D}
\ee
where $\Delta \Phi=(\Phi_{1} + \Phi_{2}) -(\Phi^{\prime}_{2} + 
\Phi^{\prime}_{2}) $ and the function
$\Phi = p^{2}_{z} - {p^{2}\over3}$ describes the deviation of the distribution
function of a collective mode from the equilibrium one. The function $W$ is
proportional to the scattering amplitude of binary collisions
$ \vec{p}_{1} \, , \vec{p}_{2} \to \vec{p}^{\prime}_{1} \, , 
\vec{p}^{\prime}_{2}$. In the vicinity of the
Feshbach resonance this function is determined by unitary limit of the
scattering amplitude and is equal to
\be
W= {\hbar^2 \over m^2} \, {(2 \pi \hbar)^3 \over p^2} \,
\label{W}
\ee
$\vec{p}=\vec{p}_{1}-\vec{p}_{2}$ is the relative momenta of two particles.

At a very low temperatures the Pauli blocking factors, $n\, (1-n)$,
in Eqs.~(\ref{N}) and (\ref{D}) significantly reduce the phase space of 
particles whose collisions appreciably contribute to the relaxation. The main 
contribution is from the collisions of particles whose
momenta lies very close to the Fermi surface:
\be 
\epsilon - \mu + V(r) \sim k_B \, T
\ee
where $\mu$ is the chemical potential Eq.~(\ref{mu}).
 
After a lengthy but a straightforward calculation one obtains for a damping
rate:
\be
\Gamma \approx {9 \pi \over 50} \, {\left(k_B \, T\right)^2 \over
\hbar \, \mu} \,.
\label{Gammagas}
\ee
It is important to note a temperature dependence of the damping rate. It has
a typical $T^2$ dependence which comes from the life time of weekly 
interacting Fermi particles or quasi-particles in the case of the Fermi
liquid \cite{L&L}. 
We stress that such scaling is true only for weekly interacting
gas of particles or quasi particles which is observed in liquid $^3\!He$.
If the atomic cloud near Feshbach resonance is indeed a strongly interacting
near perfect liquid the damping rate and other dissipative processes will
very weekly depend on the temperature. This prediction can be checked 
experimentally.

The damping rate can be expressed in terms of the trap frequencies
and the Fermi temperature $T_{F}$ as
\be
\Gamma \approx {9 \pi \over 50} \, \left(3 N\right)^{1/3} \, \bar{\omega} \,
\left({T\over T_{F}}\right)^2
\label{GT}
\ee
For the experiment of Bartenstein {\it et al.} \cite{osc_omega_damp}
the above formula predicts for the axial mode
\be
{\Gamma_{z} \over \omega_z} \approx 0.56 \,.
\ee
This is much larger than the damping observed close to the Feshbach resonance,
and is only compatible with data well away from it, where $k_F \, a < 1$.

Similarly, for the experiment of Kinast {\it et al.} \cite{duke2}
the Eq.~(\ref{GT}) predicts the ratio for the damping of the radial mode
to be
\be
{\Gamma_{r} \over \omega_r} \approx 0.17 \,,
\ee
which is almost ten times larger then the observed
ratio $\Gamma_{r}/ \omega_r \approx 0.014$. 

Thus, the strongly interacting atomic cloud is better
described by a picture of a nearly perfect strongly interacting liquid
with viscosity very close to the minimum possible value.
As one moves away from the Feshbach resonance into regime of weekly
interacting Fermi gas the kinetic theory becomes again applicable.
Furthermore, as detuning gets larger than used in the experiment discussed,
the scattering length is getting small, the gas will enter an almost
collisionless regime, and the  damping rates is getting small again.
Thus the damping rate is expected to reach a maximum value for certain value
of the magnetic field, separating weakly and strongly coupled regimes.

For the radial mode the value predicted by the same kinetic calculation is
\be
{\Gamma_{r} \over \omega_r} \approx 0.02 \,,
\ee
comparable to the value of about  $0.06$ obtained experimentally. 
It shows that this mode may not be treated by hydrodynamics.

\section{Discussion} \label{sec_discussion}

\subsection{Other strongly coupled systems} \label{sec_other}

In the introduction we mentioned a number of strongly
coupled systems which display similar transport properties.
One example is the strongly coupled quark-gluon plasma 
(sQGP) \cite{sQGP} which was found in heavy ion experiments at RHIC
at temperatures above the critical temperature, $T=(1-2)T_c=170-350\, MeV$.
In ultrarelativistic heavy ion collisions also quite spectacular 
explosions are observed, with radial and elliptic flows
surprisingly well described by ideal hydrodynamics. The sQGP seems to have  
a very small viscosity, i.e. $\eta/s\approx 0.1-0.2$ \cite{hydro,DEREK} and 
$\eta/s \approx 0.2-0.4$ from lattice simulations \cite{LAT}. 
It is not even far from the lowest limit discussed below.
Two of us suggested that the sQGP has such low viscosity because 
of the existence of weakly bound states near the so called 
``zero binding lines'' on the QCD phase diagram \cite{sQGP}. 
It was gratifying to learn after that
that similar role for trapped atoms is played by Feshbach resonances.

Other examples are: strongly coupled ${\cal N}=4$ Supersymmetric Yang-Mills
(SYM) gauge theory, a four dimensional conformal field theory (CFT) and 
the strongly coupled QED plasma. The classical one-component QED plasma 
is particularly well studied and is known 
at values of the effective coupling 
$\Gamma=(Ze)^2n^{1/3} /T =2-300$ to be a liquid,
with a very small viscosity which has its minimum  
at $\Gamma\sim 10$ (see e.g.\cite{scp}). 

The CFT is a toy model, emerged mostly in the context of
string theories. 
It is a model resembling QCD, the gauge theory of strong interaction.
In CFT as opposed to in QCD the gauge coupling is 
not ``running'' with energy scale. Thus,
the strong coupling regime is obtained simply by using a large  
coupling constant in the Lagrangian, $\lambda=g^2N$.
(Here $g$ is the gauge coupling and $N$ is the number of colors. Only this
combination appears if $N$ is large.) This limit can be addressed using
the AdS/CFT correspondence as originally suggested by 
Maldacena~\cite{maldacena}, whereby the quantum intricacies of the 
strongly coupled gauge theory are mapped onto a classical problem in 
gravity albeit in ten dimensions. The finite temperature version of this theory
describes a plasma-like phase with strongly coupled constituents. 
The four dimensional world (in which the CFT fields live) is a surface
in ten dimensional space, at some distance from a 
black hole, with a mass adjusted to yield the desired temperature $T$
at this surface. We would like to mention two important results that 
follow from this construction. One is the equation of
state of the underlying gauge theory at strong coupling 
$\lambda\gg 1$~\cite{thermo}
\be \label{EoS_CFT}
{p_\lambda (T)\over p_0(T)}= \left(1-{1\over 4}+ 
{\cal O}(1/\lambda^{3/2})\right)\,\,,\ee
where $p_0(T)\sim T^4$ is the Stephan-Boltzmann pressure for zero
coupling.
The second result is the viscosity of the underlying gauge theory 
at strong coupling \cite{PSS}
\be \label{relation_SYM}
\lim_{\lambda\rightarrow\infty}\frac {\eta}{\hbar s }=\frac {1}{4\pi}\,
\left(1+{\cal O}(1/\lambda^{3/2})\right)\,\,,
\ee
given in units of the free entropy density s\footnote{In thermal gauge
theories as in the case of blackbody radiation, there is no
ordinary particle density but only the entropy density $s\sim T^3$. }.  
The corrections were recently calculated in~\cite{Buchel:2004di}.

While the pressure is only changed by 1/4 when 
the coupling is changed from zero to infinity, the viscosity $\eta/s$
changes from infinity to a finite (and surprisingly small) number.
Thus, one may wonder whether other strongly coupled systems show
similar behavior, and whether such limiting numbers can be
universal and theoretically understood.
  
The holographic principle in the Maldacena limit and the Kubo formulae
show that the viscosity is proportional to the graviton absorption cross
section in bulk by the black hole, while (according
to Beckeinstein-Hawkins argument) the free entropy is related to its area. 
As a result  the same limit for the viscosity  holds for a number of 
backgrounds, even in different space dimensions. These observations led
to a conjecture that Eq.~(\ref{relation_SYM}) is a universal lower bound 
valid for any thermal system in strong coupling \cite{Starinets}. 
Below we provide its heuristic derivation of 
based on the uncertainty relation and Einstein's famous relation between the 
diffusion constant from the fluid viscosity. As a result, we show how 
the bound (\ref{relation_SYM}) fits well into the
the liquid-like picture of CFT at finite temperature. We then use these
insights to derive an even lower bound for cold  Fermi systems,
and conjecture that at strong
coupling they also make an universal near-ideal liquids.

\subsection{Bounds on Transport Coefficients}
 \label{sec_bounds}
 
For atomic systems one may also think that as the interaction 
strength is driven to infinity
the transport parameters such as viscosity become as small as possible.

In a weak coupling limit 
the viscosity and diffusion coefficients are both related
to the scattering length and are thought of as proportional to each other. 
For liquids one should think differently, An example of quite an
opposite relation between them was provided by the famous Einstein
relation, which we now derive for consistency of the presentation.

The distribution of suspended particles in a thermalized column of gas
is given by statistical mechanics. Indeed, if $n(x)$ is the suspension 
density at finite temperature then

\be
\frac{n(x)}{n(0)} = e^{-mgx/k_BT}
\label{1}
\ee
which follows from Boltzmann. Einstein's observed that an arbitrary
sphere of radius $r_0$ in suspension within the column would also follows
the same ``distribution'' profile. The idea is that the sphere under 
gravity will fall with a terminal Stokes velocity

\be
v_T=\frac{mg}{6\pi\,r_0\,\eta}
\label{2}
\ee
but the fall will be balanced by random upward kicks due to Brownian
motion. In equilibrium, the upward diffusion balances the downward
gravitational fall so that in the stationary limit

\be
D\,\frac{dn}{dx}= - n\,v_T
\label{3}
\ee
from which it follows that $n(x)/n(0) =e^{-v_T\,x/D}$. Comparing this
result with Eq.~(\ref{1}) and using Eq.~(\ref{2}) yields the Einstein's formulae

\be
D=\frac{k_B\,T}{6\pi\,r_0\,\eta}\,\,.
\label{4}
\ee
Although this formulae was derived for a macroscopic sphere of 
radius $r_0$ immersed in a suspension, empirically it is known to
hold through 14 orders of magnitude changes down to the suspension
constituent wavelength \cite{exp_Einstein}.

We recall that in three dimensions the diffusion constant is just 
$D=v^2\tau/3=\lambda^2/(3\tau)$ where $\lambda$ and $\tau$ are the
mean-free path and collision time 
\footnote{We note that in $p$ space-dimensions
the diffusion constant is $D=\lambda^2/p\tau$ and (\ref{4}) should
be derived accordingly. All the bounds to follow can be extended
readily to $p$ space dimensions.}. Inserting this result into 
Eq.~(\ref{4}) yields
\be
\frac{\eta}{1/(r_0\lambda^2)}=\frac 1{2\pi}\,(k_B\,T\tau)\,\,.
\label{5}
\ee

In a densely packed liquid the smallest jump (the mean-free path $\lambda$) 
is the size of the quasiparticles $r_0$.
Classically in densely packed hard balls $\tau$ can be as small as
zero due to the fact that they are always touching. Quantum mechanically
however this is not allowed since the  time localization cannot be better
than the limit set by the {\em largest allowed energy},
by the Heisenberg uncertainty
principle, i.e. $k_BT\,\tau\geq\hbar/2$. Inserting this result into 
Eq.~(\ref{5}) yields Eq.~(\ref{relation_SYM}) 
since the entropy per unit volume $s$ is just the
number of (quasi)particles per unit volume due to the close packing,
{\it i.e.}  $s/k_B=n=1/\lambda^3$. The ensuing physical picture of the 
strongly 
coupled  thermal system is that of a liquid with the shortest time correlation 
length $\tau_{min}=\hbar/2k_B\,T$. 

Our heuristic derivation of Eq.~(\ref{relation_SYM}) follow from 
the assumption that Eq.~(\ref{4}) holds for the liquid particles,
since the relation is known to hold over many orders of magnitude
changes in $\eta, D$. Thus the particle and entropy densities are
the same. While classically the collision time is zero for
the densely packed liquid, quantum mechanically it is bounded from
below by the Heisenberg uncertainty principle. Thus,
\be 
{\eta\over s} \geq {\hbar \over 4\pi\,k_B} 
\ee
which is the same as the CFT limiting value. Turning
the argument around through Eq.~(\ref{4}) implies an upper bound
on the diffusion constant in a strongly coupled liquid, namely
\be
\frac {D}{\sigma_0}\leq \frac {k_BT}{12\,h}
\label{D1}
\ee
in three dimensions, where the cross section $\sigma_0=8\pi\,r_0^2$.

Let us now turn to cold atomic gases and repeat the same argument
once more. It is simpler to imagine a Fermi gas in a vertical gravity field,
for which we will rerun Einstein's derivation. We note that the trap
field actually fulfills the same role. Using either Thomas-Fermi or
hydrostatic calculations one finds that the Fermi momentum 
for non-relativistic quasiparticles in a gravitational
field is
\be
\hbar k_F=\left(2m^*(\mu-mgx)\right)^{1/2},
\label{6}
\ee
where $m^*$ is the quasiparticle mass and $m$ is its bare gravitational mass. 
The normal Fermi density in a weak gravitational field is
\be
\frac{n(x)}{n(0)} = \left(1-\frac {mgx}{\mu}\right)^{3/2}
\approx e^{-3mgx/(2\mu)}
\label{7}
\ee
Using Eq.~(\ref{7}) and repeating steps which led to Eq.~(\ref{5})
one gets,
\be
\frac{\eta}{1/(r_0\lambda^2)}=\frac 1{3\pi}\,(\mu\tau) 
\label{8}
\ee
with a diffusion constant

\be
D=\frac{\mu}{9\pi\,r_0\eta}\,\,.
\ee
In the infinite coupling limit $\lambda$ becomes $r_0$ and the system
is again closely packed. The Heisenberg uncertainty principle stipulates
that the shortest collision time is dictated by the largest available
quasiparticle energy, namely $\mu\,\tau\geq \hbar/2$. Thus the new bound
on the viscosity
\be
\frac{\eta}{\hbar n}=\alpha_\eta \geq \frac{1}{6\pi}\,\,.
\label{NEW}
\ee

Although Eq.~(\ref{NEW}) was derived for non-relativistic particles,
its insensitivity to the quasiparticle velocities imply that it
should hold in the relativistic case as well. Equation (\ref{NEW}) 
implies an upper bound on the diffusion constant
\be
\frac D{\sigma_0}\leq \frac{\mu}{18h}\,\,,
\label{D2}
\ee
in trapped cold Fermions in three dimensions.

\section{Conclusions}

Strongly interacting systems, in relativistic field theories 
(QCD, CFL) and in condensed matter physics 
(strongly coupled QED plasma, ultracold atoms near the Feshbach resonance),
are radically different from gas-like weakly interacting systems.
The best way to see that is $not$ via the equation of state and
related thermodynamical observables, but with the help of  transport
properties.
 
As we have argued above, all such systems are near-ideal
liquids, and thus the natural tool one should use to describe those are
standard viscous hydrodynamics.
Indeed, one gets very good description of the ``elliptic flow'',
the frequency and damping of the lowest collective oscillations.

Furthermore, the viscosity extracted from the data 
are shown to be very different
from what is expected on the basis of binary collisions
and in the weak coupling regime. We compared 
the minimal experimental value of the viscosity 
(or maximal rescattering rate) Eq.~(\ref{eqn_minvalue}),
with results of standard kinetic theory.
Even without Pauli blocking and with
maximal (unitary limited) binary cross section the kinetic theory
fails to reproduce the data. 
We have also shown that the kinetic theory describes
damping far from Feshbach resonance, where the
the weakly coupled gas-like regime is valid.
We conclude that the system of optically trapped ultracold atoms
near the Feshbach resonance, like other 
strongly coupled systems discussed in section~\ref{sec_discussion},
is not a gas but rather a near-ideal liquid. Since 
oscillations are maximal at the surface, this small viscosity
does not refer to superfluidity and is property of the normal
component of the liquid making. Thus, the the properties of the
normal component of the atomic system near the Feschbach
resonance is quite different from that of liquid $^{4}He$ below the 
$\lambda$-point. 

At infinite coupling the constituents are effectively large and densely
packed. The packing fluctuates over short time scales dictated
by the Heisenberg uncertainty principle. These physical insights
together with Einstein's description of diffusion in viscous liquids,
allow for a simple derivation of the viscosity to entropy density
ratio established using CFT. We have extended this derivation to
cold Fermi systems and derived an even lower bound for the viscosity
to the particle density ratio. 

It would be quite important to make more accurate
measurements of the damping rate 
in order to see how close is the minimal viscosity
to the theoretical bound. The temperature dependence of the damping rate
would clarify the issue of Pauli
blocking. We expect that the temperature dependence
should be weak and not even close to what binary collision theory for Fermi gas
predicts: the system is not even qualitatively
close to a Fermi gas at the strong coupling limit. 

Temperature dependence 
is also crucial for understanding of the transition to superfluidity
which was avoided in this paper. We argued that in super-component
resigns in the interior of the system, which is not very important
for elliptic flow or oscillations we discussed. 
It seems that the
lowest oscillation mode is well described by the one-component 
hydrodynamics: however it is of course quite likely that 
some higher excitations are 
analogous not to the usual sound but to other sounds known for 
superfluid liquid He.

One should also develop a theory of the strongly
coupled systems beyond quite schematic mixtures of the 
BCS superconductor and ideal Bose gas of molecules. So far,
such models can approximately reproduce the
equation of state but not the transport properties.  
Numerical simulations of larger scales can also be helpful: 
perhaps one should complement the equation of state calculations by 
measurements of long-time correlators related by Kubo relations to transport
coefficients.

\vskip 0.5cm

{\bf ACKNOWLEDGMENTS}

\vskip 0.5cm

We thank Gerry Brown for numerous and very helpful  discussions
on Fermi liquids.
We also thank H.T.C.~Stoof for a useful discussion at the early stage of 
this work and R.~Grimm. We especially grateful to 
J.E.~Thomas for helpful correspondence.
This work was partially supported by the US-DOE grants DE-FG02-88ER40388
and DE-FG03-97ER4014.

\end{narrowtext}
\end{document}